\documentclass[twocolumn,pre,tightenlines,superscriptaddress,showpacs]{revtex4-2}

\usepackage{amsmath}
\usepackage{amssymb,amsfonts,latexsym}
\usepackage{bm}
\usepackage[mathcal]{euscript}
\usepackage{graphicx}
\usepackage{epsfig}
\usepackage{color}
\usepackage{xfrac}
\usepackage{mathrsfs}

\usepackage{hyperref}
\usepackage{url}
\usepackage{CJK}

\begin{document}

\begin{CJK*}{UTF8}{gbsn}

\title{Coexistence phase of the two-dimensional Vicsek model}

\author{Yu Duan ({\CJKfamily{gbsn}段煜})}
\affiliation{Center for Soft Condensed Matter Physics and Interdisciplinary Research \& School of Physical Science and Technology, Soochow University, Suzhou 215006, China}
\affiliation{Max Planck Institute for Dynamics and Self-Organization (MPI-DS), 37077 G\"ottingen, Germany}
\affiliation{Jiangsu Physical Science Research Center, Nanjing 210093, China}

\author{Hugues Chat\'{e}}
\affiliation{Service de Physique de l'Etat Condens\'e, CEA, CNRS Universit\'e Paris-Saclay, CEA-Saclay, 91191 Gif-sur-Yvette, France}
\affiliation{Sorbonne Universit\'e, CNRS, Laboratoire de Physique Th\'eorique de la Mati\`ere Condens\'ee, 75005 Paris, France}

\author{Beno\^it Mahault}
\affiliation{Laboratoire Charles Coulomb, Universit\'e de Montpellier, CNRS, Montpellier, France}
\affiliation{Max Planck Institute for Dynamics and Self-Organization (MPI-DS), 37077 G\"ottingen, Germany}

\author{Xia-qing Shi ({\CJKfamily{gbsn}施夏清})}
\affiliation{Center for Soft Condensed Matter Physics and Interdisciplinary Research \& School of Physical Science and Technology, Soochow University, Suzhou 215006, China}

\date{\today}

\begin{abstract}
The Vicsek model continues to be central in statistical physics studies of active matter, with major recent surprises such as the partial metastability of the famous Toner-Tu flocking phase. Here we report on the diversity of band patterns observed in its coexistence phase --some of them never observed before--, but we also argue that most disappear in the infinite-size limit, leaving nevertheless a different picture than anticipated previously. 
We finally show that these rather regular, globally ordered band configurations can be destroyed by a localized perturbation: the coexistence phase is also at least partially metastable, and thus must be, asymptotically, a disordered phase made of a variety of chaotically interacting traveling bands. 
\end{abstract}

\maketitle
\end{CJK*}

\section{Introduction}

Active matter often designates collectives where at least a fraction of the composing units transduce energy, 
either stored internally or gathered from the environment, into mechanical work~\cite{ramaswamy2010mechanics,marchetti2013hydrodynamics,bechinger2016active,tevrugtColloquiumWhatWe2026}. 
This mechanical work can be used in various ways, among which self-propulsion of the active units is probably
one considered the most.

Very common among such self-propelled particles used in the simplest models of active matter, are active Brownian particles~\cite{romanczuk2012active}. 
These are usually taken to be persistent random walkers moving at constant speed $v_0$. Their position ${\bf r}$ and velocity orientation $\theta$ are typically governed, in two space dimensions, by the following equations:
\begin{equation}
\dot{\bf r} = v_0 \,{\bf e}(\theta) \;\;\;\; \dot{\theta} = \sqrt{2D_r} \, \xi
\end{equation}
where  ${\bf e}(\theta)$ is the unit vector of orientation $\theta$ and $\xi$ is a zero-mean unit-variance Gaussian white noise.

Models consisting only of interacting self-propelled particles neglect the fluid that usually surrounds 
realistic active units. 
One then speaks of dry active matter~\cite{toner2005hydrodynamics}. The simplest and best studied models of dry active matter are made 
of active Brownian particles
interacting locally either solely via pairwise repulsion~\cite{fily2012athermal,redner2013structure} or solely via alignment of their orientations. 
The first case is well known to be able to lead to motility-induced phase separation, perhaps the most studied phenomenon
in scalar active matter~\cite{cates2015motility}. 

The second case, which historically precedes the first one, is best represented by the Vicsek model~\cite{vicsek1995novel}, in which 
active Brownian particle $j$ aligns ferromagnetically its velocity with that of its neighbors $k$:
\begin{equation}
\label{VMcont}
\dot{\bf r}_j = v_0 \,{\bf e}(\theta_j) \;\;\;\; \dot{\theta_j} = \langle \sin(\theta_k - \theta_j) \rangle_k + \sqrt{2D_r} \, \xi_j \;,
\end{equation}
where we have, in all generality, taken a unit coupling constant in front of the interaction term, 
and defined the neighbors $k$ to be particles within unit distance of $j$. 

The above equations define one particular instance of the Vicsek model. 
Many variants have been considered~\cite{czirok1996formation,chate2008modeling,ginelli2016physics}.
Notably, many works consider discrete time steps~\cite{vicsek1995novel,gregoire2004onset}, as in
\begin{equation}
\label{VMdisc}
{\bf r}_j^{t+1} \!= \mathbf{r}_j^t + v_0 \,{\bf e}(\theta_j^{t+1}) \;\;\;\;  \theta_j^{t+1} \!= \arg[\langle \exp(i\theta_k^t)\rangle_k] + \eta \, \xi_j
\end{equation}
where the average now includes $j$ and $\xi_j$ is drawn uniformly in the interval $[-\pi;\pi]$. 

Both these versions are used for the results presented below. They both display the same phenomenology. Many other versions also share these collective properties~\cite{ginelli2016physics,chate2020dry,martin2021fluctuation}.
(Note, in particular, that in the original 1995 paper~\cite{vicsek1995novel}, Vicsek {\it et al.} 
use a time-stepping scheme different from that of \eqref{VMdisc}.)

Given their simplicity, standard Vicsek models such as those defined above possess only two parameters, 
the particle speed $v_0$ and the noise strength, to which one must add $\rho_0$, the number density of particles. 
So far, $v_0$ has not been shown convincingly to influence qualitatively the collective properties of the VM~\cite{chate2008collective}, and 
the usual phase diagram is thus in the (density, noise) plane~\cite{solon2015phase}. 
Below, we vary these two main parameters and perform simulations in square or rectangular domains 
$L_x\times L_y$  with periodic boundary conditions. 
For the continuous-time model defined by \eqref{VMcont} we use an explicit Euler-Maruyama scheme~\cite{higham2001algorithmic} with typical timestep $dt=0.1$. 

As we explain below, numerous open questions and new facts have emerged over the years that question our 
current knowledge of the two-dimensional Vicsek model. This has led us to explore afresh its collective properties 
on timescales and lengthscales almost never probed before. The gist of this mostly numerical campaign is
summarized in a short paper (SP) \cite{SP} which presents the resulting updated phase diagram, 
while here we give an account of the wealth of results obtained in the coexistence phase of the model. 
Theoretical aspects are mostly absent, although of course our current theoretical knowledge of the model 
has informed our work, and some theoretical consequences of our findings are discussed at the end of both papers.  

\subsection{Previous knowledge}

Under the form of Eqs.\eqref{VMcont}, the two-dimensional Vicsek model appears to be an off-lattice 
XY model in which the 
spins move in space at constant speed along their orientation $\theta$, interacting with whoever is 
within unit distance of them. 
Naively, one would expect a phenomenology similar to that of the XY model~\cite{kosterlitz2017nobel}: a ferromagnetic quasi long-range ordered phase 
at low noise when the spins are globally aligned (and thus flocking collectively) and a disordered paramagnetic phase for strong enough noise. 
This was indeed the viewpoint of Vicsek and collaborators when they introduced the model \cite{vicsek1995novel}.

But there are (at least) two fundamental differences between the VM and the equilibrium XY model. 

First, as predicted early by Toner and Tu (TT) \cite{toner1995long}, 
the homogeneous flocking phase possesses true long-range polar order \cite{mahault2019TT,chate2020dry},  
something prevented in equilibrium by the Hohenberg-Mermin-Wagner theorem \cite{mermin1966absence,hohenberg1967existence,tasaki2020hohenberg}, 
and it is metastable in part of its domain of existence \cite{codina2022small,granek2026how}.

\begin{figure}[t!]
  \includegraphics[width=\columnwidth]{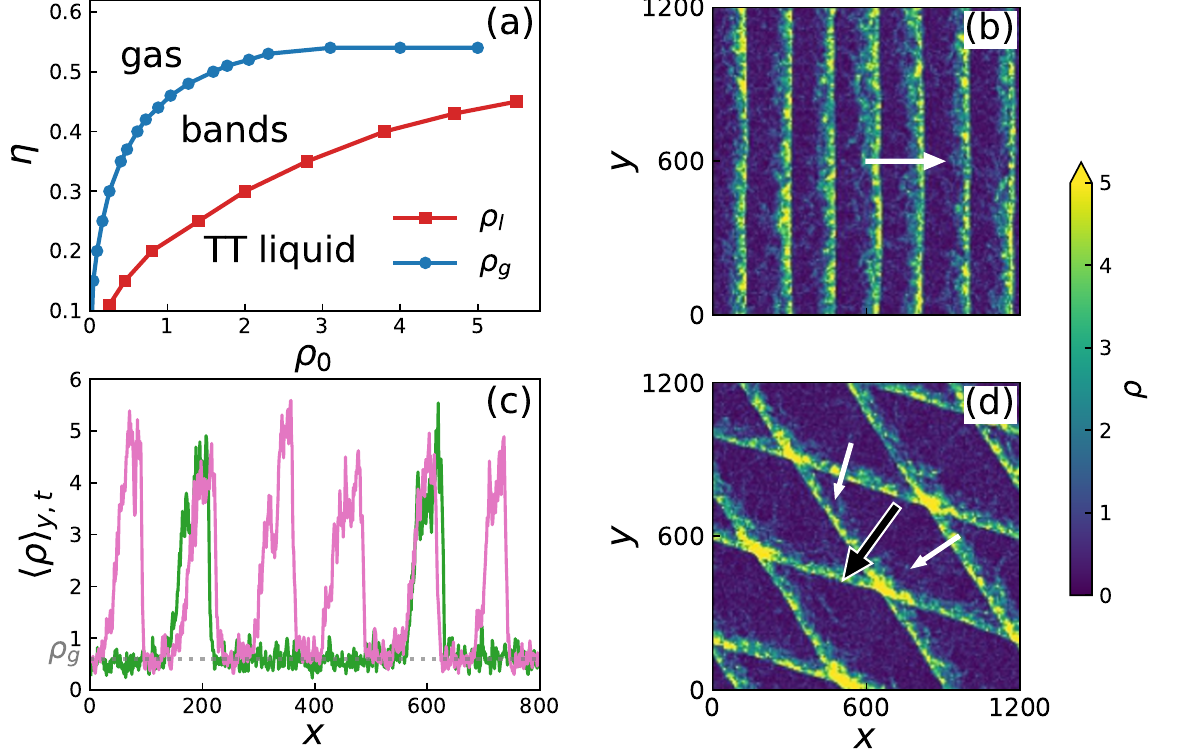}
    \caption{Current knowledge about the coexistence phase of the Vicsek model. 
    (Data taken from \cite{solon2015phase} for (a,c).)
    Results from the discrete-time model \eqref{VMdisc} with $v_0=0.5$.
    (a) phase diagram in the (global density, noise strength) parameter plane. 
    (b) typical coarse-grained density field of a smectic arrangement of bands ($L_x = L_y = 1200$, $\eta=0.32$, $\rho_0=1$, coarse-grained over boxes of linear size 2, the white arrow indicates the direction along which bands travel).
    (c) superposition of instantaneous density profiles (averaged over the transverse direction) in smectic 
    configurations at two different global densities (bands traveling from left to right,
    $\eta=0.4$, $\rho_0=1.05$ (green line) and $\rho_0=1.93$ (pink line); the grey dashed line indicates the gas density $\rho_g$).
    (d) same parameters as (b) but showing a stable ``cross-sea'' configuration of bands. The large black arrow indicates the 
    orientation of global polar order.
 }
\label{fig1}
\end{figure}

Second, there is no direct order-disorder transition 
in the VM and other dry aligning active matter systems \cite{gregoire2004onset,chate2008collective,chate2020dry}.
It has been shown that the correct description is instead in terms of phase separation \cite{solon2015phase}:
the homogeneous ordered flocking ``Toner-Tu liquid"  is separated from the disordered ``gas" by a coexistence phase
where particles self-organize into high-density high-order bands traveling in the remaining sparse gas 
(Fig.~\ref{fig1}(a)). The two transition lines bordering the coexistence phase in the standard 
(global density, noise strength) parameter plane
can thus be called the gas and the liquid binodals.
The traveling bands are microphases, to be understood as quanta of the ordered collectively moving TT liquid. 
They have been shown to often arrange themselves in smectic configurations with well-defined  
density and order profiles depending only on the noise strength
(Fig.~\ref{fig1}(b,c)). Their number is then proportional to
the system size or the global density of particles~\cite{solon2015phase}.

\subsection{Pending issues}

Even though the above results are backed up by a reasonably thorough theoretical understanding offered by 
continuous descriptions~\cite{bertin2009hydrodynamic,mishra2010fluctuations,gopinath2012dynamical,ihle2013invasion,solon2015pattern}, 
important pending issues remain, notably regarding the coexistence phase, 
which we now list before addressing them in the rest of this paper.

\begin{itemize}
\item The smectic arrangements of traveling bands such as that shown in Fig.~\ref{fig1}(b)
are observed in rectangular geometries of modest dimensions and at relatively weak noise.
Bands tend to become wider when increasing noise, making them more difficult to study, 
all the more so since the 
two binodal lines get closer and closer to each other when $\rho_0$ increases (Fig.~\ref{fig1}(a)). 
A thorough investigation of the coexistence phase at larg-ish densities is missing.

\item Approaching the liquid binodal line implies fitting, at a given domain geometry, more and more bands, so that there may not be enough space to accommodate them in a smectic arrangement. One then often observes two trains of bands crossing each other at some angle, which allows to fit a longer total band length (Fig.~\ref{fig1}(d)). 
Such configurations, which had been spotted in \cite{chate2020dry},
have been coined ``cross-sea''. 
But bands crossing each other were also reported in a large fraction of the coexistence region 
(also in \cite{kursten2020dry}). It is thus not clear whether cross-sea configurations 
actually constitute a separate sub-phase 
of the coexistence region in \cite{kursten2020dry}.

\begin{figure}[t!]
  \includegraphics[width=\columnwidth]{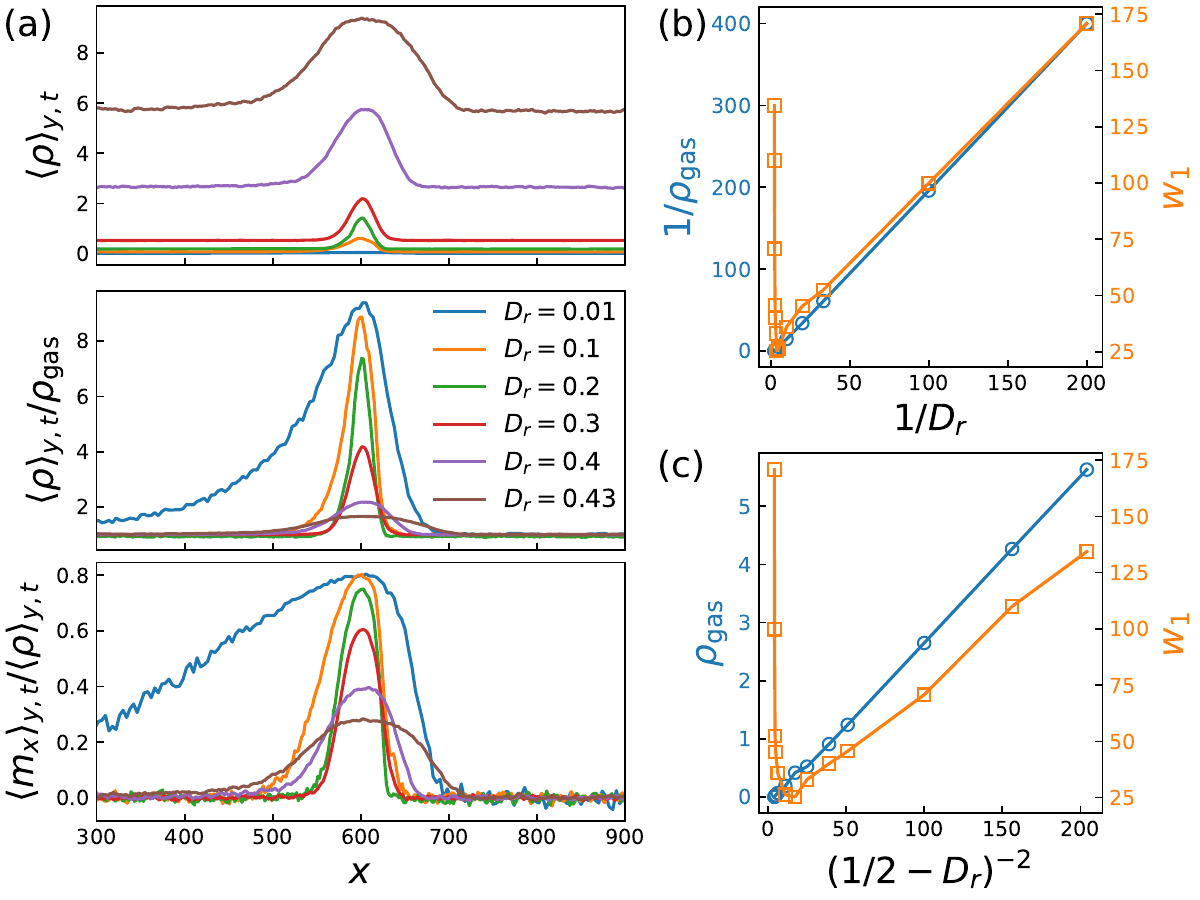}
    \caption{Scaling of the gas binodal and the first band emerging near it in the vanishing noise and infinite density limits. The band is moving along the $+x$ direction (results obtained with the continuous-time VM \eqref{VMcont} with $v_0=1$, $L_x=1200$, $L_y=300$).
    (a) part of time- and $y$-averaged profiles at various noise strengths (density, density relative to the gas density, and polar order in, respectively, the top, middle, and bottom panels).
    (b) $1/\rho_{gas}$ (blue data, left $y$-axis) and $w_1$ (orange data, right $y$-axis) vs $1/D_r$ revealing the scaling of these quantities in the zero-density zero noise limit.
    (c) $\rho_{gas}$ (blue data, left $y$-axis) and $w_1$ (orange data, right $y$-axis) vs $(\tfrac{1}{2}-D_r)^{-2}$ revealing the scaling of these quantities in the infinite-density limit.
 }
\label{fig_two_lim}
\end{figure}

\begin{figure*}[t]
  \includegraphics[width=\textwidth]{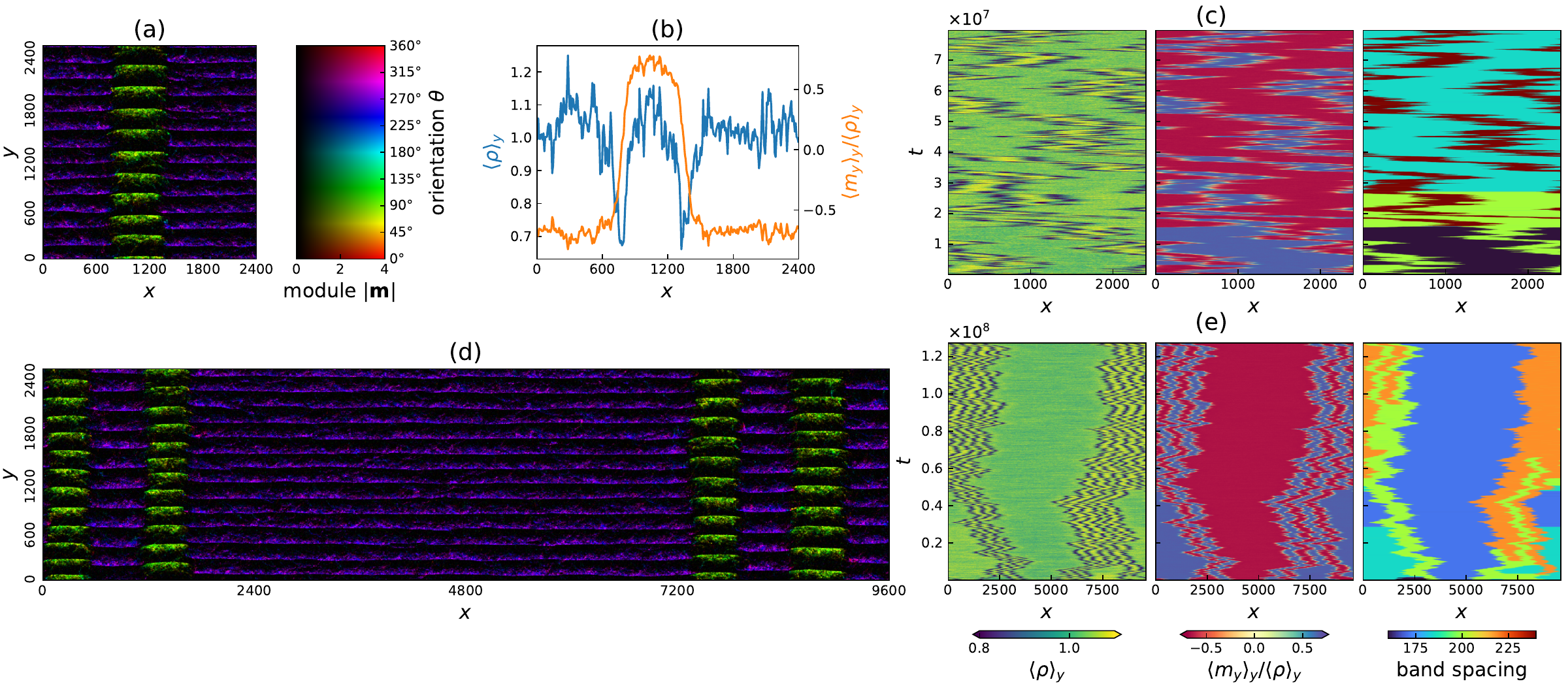}
    \caption{Counter-going band lanes. Results obtained with the discrete-time model \eqref{VMdisc} with $v_0=0.5$, $\rho_0=1$, $\eta=0.29$.
    (a) momentum field $\mathbf{m}(\mathbf{r})=\sum_j \delta(\mathbf{r}_j-\mathbf{r})\mathbf{e}(\theta_j)$, coarse-grained over boxes of linear size 2, of a simple two-lane configuration (10 up-going bands in the central lane, 13 down-going bands in the other wide lane) and colormap used throughout this paper to represent momentum fields. 
    (b) instantaneous density and velocity profiles (averaged along the lanes) of the configuration shown in (a).
    (c) space-time plots of density (left), velocity (centre), and band spacing profiles (right) leading to the configuration shown in (a) after $8\times 10^7$ timesteps. Note the slow diffusion of lanes.
    (d) momentum field of a wide system with narrow lane bunching, initialized from a configuration of eight equally spaced domains, with particles randomly positioned but identically oriented along either (+y) or (-y) within each domain, and alternating orientations between adjacent domains.
    (e) same as (c) but for the evolution leading to the configuration shown in (d) after $1.2\times 10^8$ timesteps.
 }
\label{fig_short_lane}
\end{figure*}

\item Whereas the gas binodal is easily and precisely located (by measuring the gas density in configurations with well-separated bands), the location of the liquid binodal remains uncertain.
Previous work, e.g. \cite{chate2008collective,solon2015phase},
have defined it in a somewhat ad-hoc manner. All this calls for clarification. 

\item Smectic and cross-sea configurations such as those in Fig.~\ref{fig1}(b,d) possess rather strong global polar order.
But the stability of very long bands (along the direction transverse to global order) is unknown, 
even in continuous theories, and thus it is actually unknown whether there is true long-range polar order 
in the infinite-size limit. 

\item Simulations of a Boltzmann equation derived from a Vicsek model \cite{thuroff2014numerical} have found 
intriguing solutions, made of two counter-going lanes of bands, that have never been observed at particle level.

\item The results presented in \cite{codina2022small} indicate that the TT liquid phase is probably metastable in a rather large part of its parameter domain of existence. The metastability of the liquid, but also that of ordered band configurations could be investigated more systematically.
\end{itemize}

The rest of this paper is organized as follows. 
In Section~\ref{twolimits} we distinguish two regimes, weak and strong noise, and discuss two limits, 
vanishing noise and infinite density, that will be useful for describing the complexity of solutions observed in the coexistence phase.
In Section~\ref{diversity}, we provide an organized account of the many band configurations observed at moderate system sizes, which include lanes of counter-going trains of bands, fat and inverse bands, and configurations with crossing bands.
Section~\ref{diagram} provides more information about how the new phase diagram presented in the SP was established. We also present there evidence that at least some of the band patterns are metastable.
Section~\ref{discussion}, after a summary of results, lists remaining important open questions 
before a brief discussion of how well known continuous theories can account for the phenomena presented here.

\section{Two limits, two regimes}
\label{twolimits}

We start by defining a weak noise and a strong noise regime, an approximate but useful distinction 
around which we organize the presentation of our results.

Near the gas binodal line, it is relatively easy to obtain configurations with a single band, from which one can readily
read the gas density $\rho_{gas}$  defining the gas binodal.
 
As reported in the SP (cf. Fig.~2(a) in \cite{SP}), $w_1$, the width of this ``first band" varies non-monotonously
with the noise strength. It diverges in both the zero-noise zero-density limit and in the 
infinite-density, global coupling limit, taking a minimum value at some intermediate noise strength.
These results, and those presented below, were obtained with the continuous-time version of the VM 
(Eq.~\eqref{VMcont} with $v_0=1$, for which $w_1$ is minimum when $D_r\simeq 0.23$). 
(Similar results were obtained in the discrete-time version of the VM (Eq.~\eqref{VMdisc}) with $v_0=0.5$, 
a case for which we find the thinnest first band for $\eta\simeq 0.33$).
To regularize the results, we followed a well-defined protocol: at each noise strength $D_r$, we find a 
configuration with only one band that remains steady during $10^6$ time-units, and then decrease the density by 0.004,
check whether it survives the same amount of time. The results below are measured from the last surviving bands.

In Fig.~\ref{fig_two_lim}(a), we show how the time-averaged profiles of the first band change along the gas binodal.
In the raw density profiles (top panel), the first band seems to get wider and more massive as the noise strength 
increases, but, relatively to the gas density (middle panel), $w_1$ is seen to first decrease, then increase again at strong noise. Similarly if not more clearly, the profiles of polar order (bottom panel) also show this non-monotonous 
variation of $w_1$.

The weak and strong noise limits show rather clear scaling behavior, which we detail here for the 
continuous-time VM before proceeding to the rest of the paper. 
In the zero-noise limit, 
consistent with earlier studies~\cite{chate2008collective, rubio2019self},
 the gas density (defining the gas binodal) is proportional to the noise variance, $\rho_{gas}\propto D_r$, 
 and the first band width diverges as $w_1 \sim 1/D_r$ (Fig.~\ref{fig_two_lim}(b)).
  In the infinite-density, strong noise limit $D_r\to\tfrac{1}{2}$, we see that the gas binodal 
 and the width of the first band diverge similarly: $\rho_{gas}, w_1\sim1/(\tfrac{1}{2}-D_r)^2$ (Fig.~\ref{fig_two_lim}(c)).

As shown in the SP and as we will see below, the intermediate noise strength at which $w_1$ is minimal 
roughly delimits a weak-noise and a strong-noise regime, with important differences in the types of solutions
and dynamics found between the two in the coexistence region.

\section{Diversity of band patterns}
\label{diversity}

Most previous studies have actually been performed in the weak noise regime,
where the familiar smectic and cross-sea band configurations have been reported. 

One can easily observe other band patterns when 
considering stronger noise and spatial domains larger than those used so far.  
We report here on those we uncovered, that fall into two types,
counter-going lanes each composed of a smectic train of bands found at intermediate noise strength, 
and systems of fat and inverse bands present in the strong noise regime.

\subsection{Counter-going smectic lanes}
\label{lanes}

In this section, we only present results obtained with the discrete time VM defined by Eqs.\eqref{VMdisc}
using $v_0=0.5$. We have checked, though, that the same phenomena occur at other speeds 
and with the continuous-time VM.

At intermediate noise strengths ($\eta=0.28-0.35$)
one can observe, after some transient evolution, counter-going lanes made of parallel thin bands similar to those 
found in the coexisting
smectic configurations. 
A typical case is shown in Fig.~\ref{fig_short_lane}(a,b), where two trains of bands going in opposite directions have 
formed~\footnote{Note that this pattern is very similar to that reported at kinetic level in \cite{thuroff2014numerical}.}.

This was actually reached from random initial conditions, but this is a relatively rare event. 
Out of 24 runs performed, only one led to a smectic lane configuration. (The others led to smectic or various crossing band configurations.) We noticed, however, that lanes become 
much more frequent when considering rectangular domains and/or preparing ``lane initial conditions" in which particles have random positions but identical orientations parallel to predefined parallel domains, alternating from one domain to the next.

\begin{figure}[t!]
  \includegraphics[width=\columnwidth]{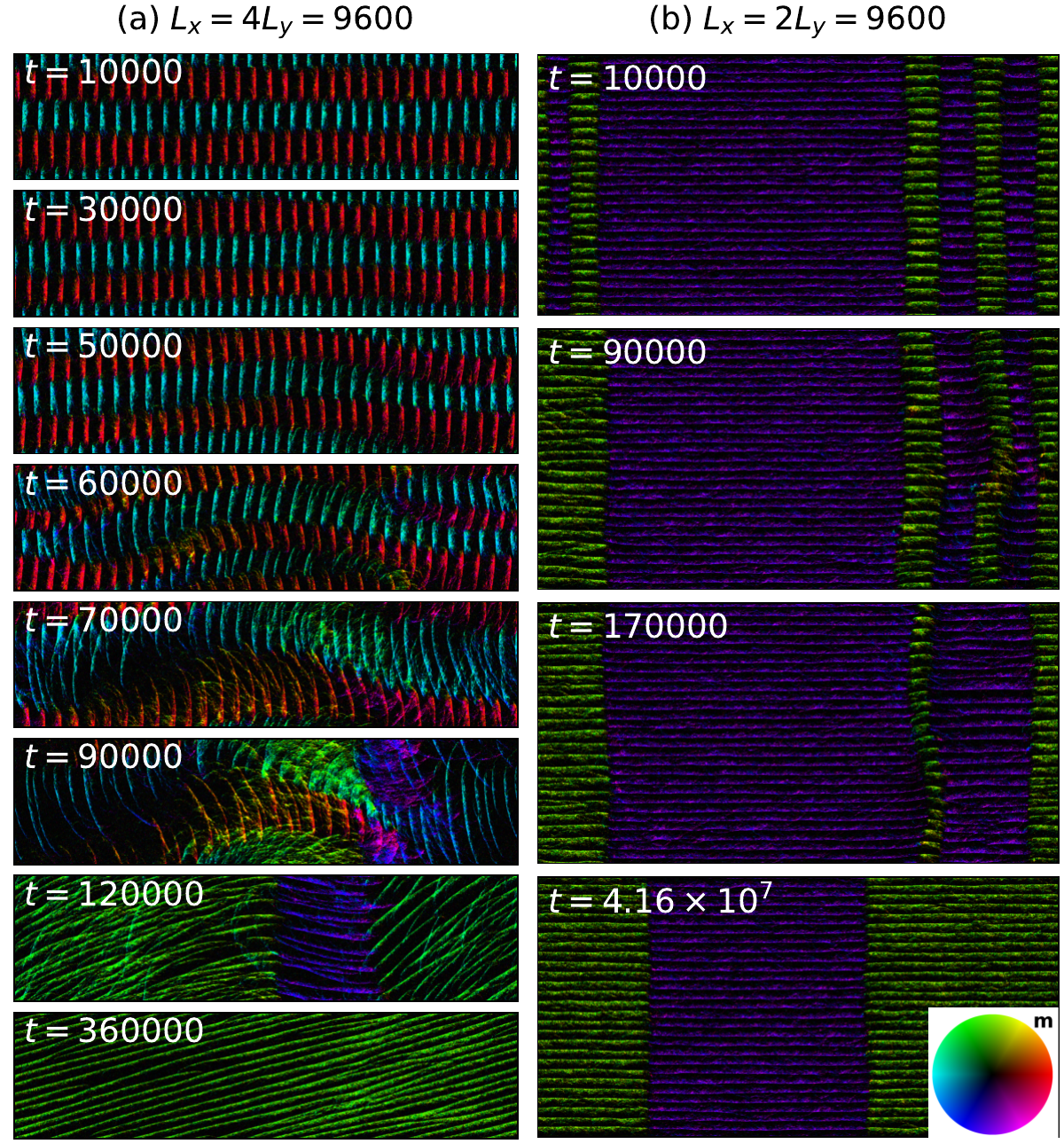}
    \caption{Instability of long lanes (discrete-time VM with $v_0=0.5$).
    Momentum field shown at various times (time running downward)
    (a) $\eta=0.35$, $\rho_0=1.264$.
    (b) $\eta=0.29$, $\rho_0=1$.
 }
\label{fig_long_lane}
\end{figure}

Irrespective of the type of initial conditions, soon after lanes appear their number stays fixed thereafter, even if their positions, their widths and the band spacing in each of them can still evolve on very long timescales (Fig.~\ref{fig_short_lane}(c)). 
In systems larger than the one shown in Fig.~\ref{fig_short_lane}(a-c), configurations with more than two lanes exist and coexist, 
see an example in Fig.~\ref{fig_short_lane}(d-e) \footnote{This is in contrast with laning in self-propelled rods, where 
coarsening of lanes proceeds \cite{shi2018self}.}

As apparent in the examples shown here, lanes typically have different widths and different band spacings.
Remarkably, the seven narrow lanes present in Fig.~\ref{fig_short_lane}(d-e) and in the Supplemental Movie (SMov)1~\cite{SUPP}
share more or less the same width and band spacing, and they `bunched' against each other after a rather long transient.
Different runs performed on the same system with different initial conditions can yield different numbers of lanes, but
in the long time limit they all share the same width $w^*$ and we expect to eventually see them adjacent to each other, leaving 
a  `remaining', typically wide, lane. This wide lane
 displays the smallest band spacing, found to be of the order of the spacing observed in the coexisting smectic configurations.

Even though smectic lanes can be observed for, it seems, as long as we wish (note that we simulated some systems for up to $10^8$ timesteps), 
at the same parameter values, regular smectic band trains can also persist for similarly long times.
As of now, we cannot quantify which is `more stable' than the other, but we believe the simpler smectics are. 
Indeed, it seems a priori more likely to observe a fluctuation leading to the destruction of a narrow lane than one leading 
to the emergence of a counter-going lane from the regular smectic. For all practical purposes, though, the involved
probabilities must be minuscule, since we never observed such events. 

All the lane configurations described above comprise lanes whose length is not excessively larger than their width. 
In rectangular domains, most initial conditions lead to bands or smectic lanes propagating along the smaller dimension. 
One can however engineer configurations with very long lanes, e.g. by collating copies of smaller existing ones. 
We observed that when such lanes are long-enough they exhibit a stationary 
instability in the long longitudinal direction, under which the lanes undergo large-scale buckling, 
something that leads to the destruction of the initial lane pattern. Two examples are shown in Fig.~\ref{fig_long_lane}.

In the first case (Fig.~\ref{fig_long_lane}(a)), the initial lane pattern will eventually be completely destroyed, 
something that should lead to either some smectic arrangement of bands, 
or to a lane configuration with bands moving along the small dimension of the domain.
In the second example (Fig.~\ref{fig_long_lane}(b) and SMov2~\cite{SUPP}), the initial narrow lanes of width $w^*$ gradually merge into a wide lane,
leaving two wide lanes, one of which should normally shrink slowly to a width $w^*$, 
but would probably become itself unstable before.

The data available currently do not allow for a quantitative analysis of what sets the wavelength of the 
buckling instability of lanes. It seems to be increasing with the width of the lane, but this cannot be a satisfactory
argument since the instability is that of the {\it interfaces} between two lanes. We also noted, though, that
the spacing between bands in a given lane tends to be larger for narrower lanes. 
This spacing is a local quantity that differs on both sides of lane interfaces. We believe it governs the buckling instability,
together with basic parameters.

All in all, extrapolating from our existing partial results, smectic lanes seem always longitudinally unstable in large enough systems. The resulting dynamics may be a long, possibly never-ending chaos of bands and lanes or, more likely,
one of the underlying regular smectic or cross-sea configurations.

\subsection{Fat bands and inverse bands}

We now report about the band configurations present in the strong noise regime.
The results below were obtained using the continuous-time version, Eqs.~\eqref{VMcont}, of the VM. 
Similar phenomena were observed with the discrete time stepping. 
In this section we fix $v_0=1$. (We did not see qualitative differences at other $v_0$ values,  but much smaller or much larger speeds would require further exploration.)

\begin{figure}[t!]
  \includegraphics[width=\columnwidth]{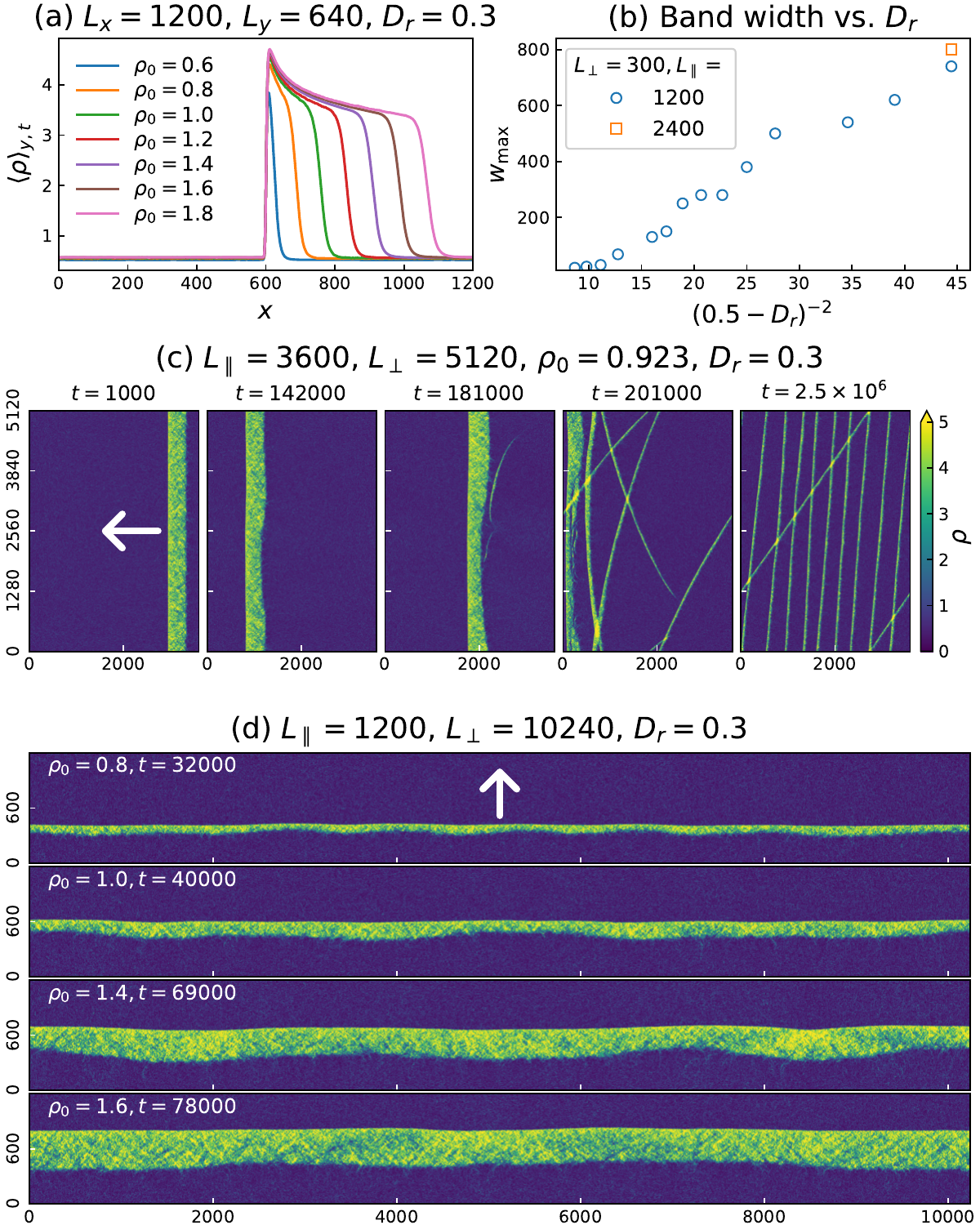}
    \caption{Fat bands.
    Results obtained with the continuous-time model \eqref{VMcont} with $v_0=1$.
        (a) superimposed time-averaged density profiles of stable fat bands obtained at various global densities. Bands are traveling to the left.
        (b) maximum width $w_{max}$ of single fat bands vs $D_r$ (obtained by increasing progressively
        $\rho_0$ until the band splits within the typical time window $\Delta t\sim 10^6$); $w_{max}$ estimated from the width at mid height of density profiles such as those in (a).
        (c) density fields at selected times showing the progressive destruction of a long fat band ($\rho_0=0.923$). The initial condition was created by duplicating a narrow , stable, single fat band traveling to the left ($L_x=3600$, $L_y=640$).
        (d) initial stages of the modulation of the trailing edge of initially straight very long upward traveling fat bands, as shown by their density fields. The long wavelength of this instability increases with $\rho_0$.
         }
\label{figFBs}
\end{figure}

\subsubsection{Fat bands}
\label{FB}

In the high-noise regime, increasing $\rho_0$ gradually at fixed noise from a configuration containing a single band, 
we do {\it not} observe the multiplication of bands characteristic of the micro-phase separation 
scenario well-known at intermediate noise. 
Instead, the initial single band gets wider and wider, developing a large homogeneous
central region, which can be hundreds of times larger than the unit interaction length,
all this while the density in the surrounding gas remains constant (Fig.~\ref{figFBs}(a) and Fig.~2(b,c) in the SP \cite{SP}).

\begin{figure*}[t!]
  \includegraphics[width=\textwidth]{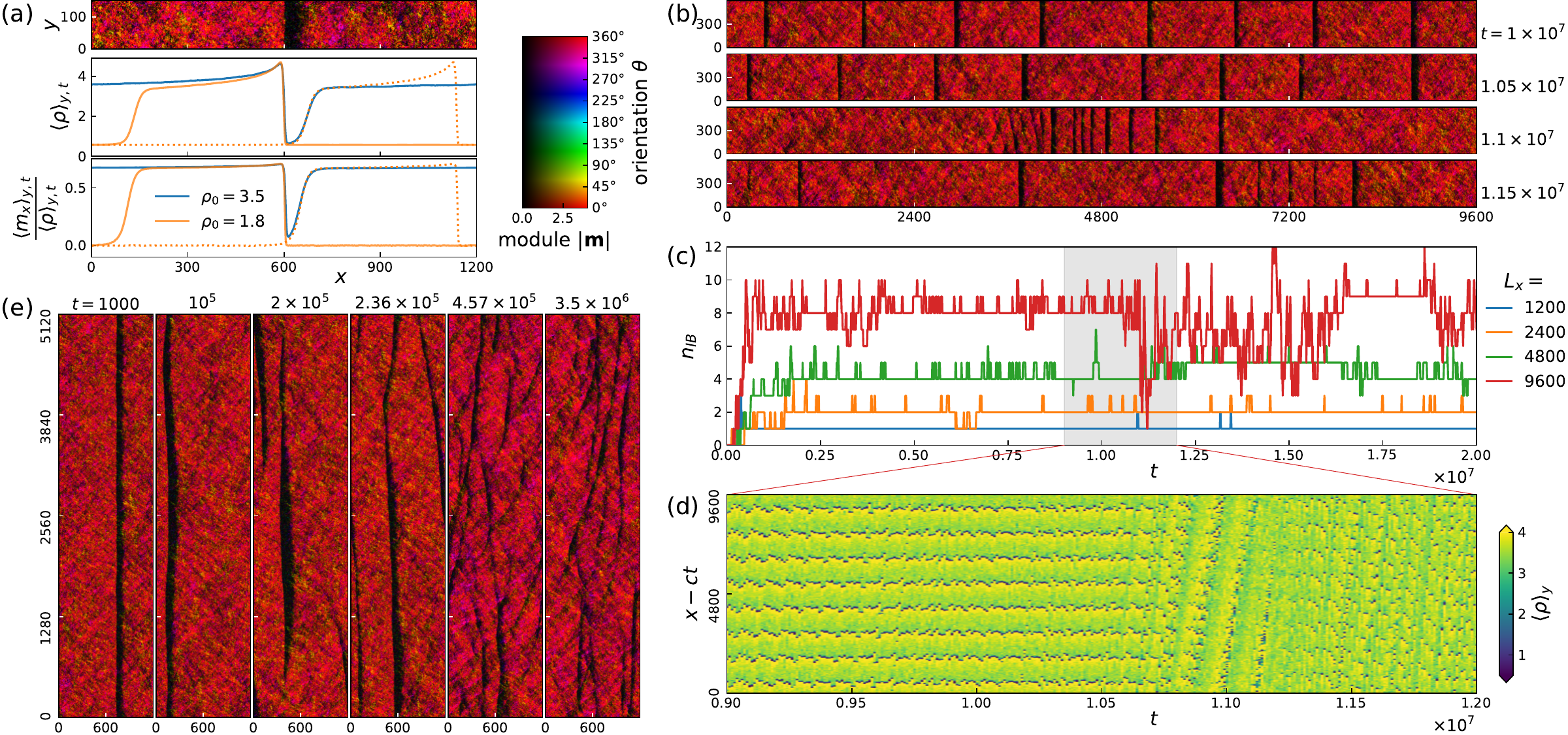}
    \caption{Inverse bands.
        Results obtained with the continuous-time model \eqref{VMcont} with $v_0=1$,  $D_r=0.3$, $\rho_0=3.5$.
        (a) stable configuration with a single, stable, inverse band ($L_x=1200$, $L_y=640$). 
        Top panel shows a snapshot of part of the momentum field, while the bottom two panels show time-averaged and $\perp$-averaged density and order profiles along the order direction (in blue), plus corresponding profiles of a fat band
        observed at the same noise strength but for $\rho_0=1.8$ (in orange), suitably shifted along $x$
        so that its leading and trailing edges ``match" the inverse band profile.
        (b) snapshots of momentum field taken at times separated by $5\times 10^5$ time units in a long system 
        ($L_\|=9600$, $L_\perp=600$)
        (c) time series of the number of inverse bands in systems of length $L_\|=1200$, 2400, 4800, and 9600, containing on average 1, 2, 4, and 8 inverse bands ($L_\perp=600$ and parameters as in (b)). Inverse bands
        were counted from the density profiles shown in (d) thresholded at $\rho=2$. 
        (d) space-time plot, in the frame moving at the speed $c=0.686$ of the default 8-inverse band configuration, 
        of the $\perp$-averaged density profile of the long system shown in (b) and (c) (red line). The time-interval represented is shown in grey in (c).
        (e) momentum fields at various times showing the instability of a long, initially straight, inverse band, created by duplicating a configuration similar to that shown in (a).
}
\label{figIBs}
\end{figure*}

This smacks of macrophase separation, where a single liquid domain is present. 
Indeed as shown in Fig.~2(d) of the SP, the lever rule is satisfied: the width of these fat bands, 
measured at mid-height from profiles such as the one shown in Fig.~\ref{figFBs}(a), varies linearly with $\rho_0$.
However, increasing the global density even further, one eventually observes the splitting of the heretofore single
fat band into two or more, whose widths then equilibrate in time.
The precise global density at which this happens seems to depend slightly on parameters and 
on the protocol used to increase $\rho_0$.
For a given such protocol, we observe that the maximal width $w_{max}$ 
for observing a single fat band increases fast with $D_r$, diverging, like the first-band width $w_1$, 
as $1/(\tfrac{1}{2}-D_r)^2$ (Fig.~\ref{figFBs}(b)). 
Increasing $L_\parallel$ does not change much to the above picture, except for a small increase of $w_{max}$ 
(cf orange square in Fig.~\ref{figFBs}(b)). 

Unless this maximal width diverges beyond a finite $D_r$ value ---something we are currently unable to probe---,
 its existence invalidates the macrophase separation scenario. 
Moreover, at the same parameter where we obtain a single fat band, other initial conditions and protocols can yield multi-fat-band configurations that we can observe on equally long timescales. 
This is further at odds with standard macrophase separation.

In any case, how $w_{max}$ varies does not matter much: for very large $L_\perp$, we observe
the breakdown of long fat bands. Illustrated in Fig.~\ref{figFBs}(c) and in SMov3~\cite{SUPP}, it typically starts
by large scale modulations of the trailing edge of the band, leading to the shedding of standard thin bands, 
resulting finally in some smectic or cross-sea configurations of thin bands. 
A careful investigation of even longer fat bands with various initial widths reveals that their trailing edges 
undergo a stationary instability with a long wavelength that increases with $\rho_0$ (Fig.~\ref{figFBs}(d)).

\begin{figure*}[t!]
    \includegraphics[width=\textwidth]{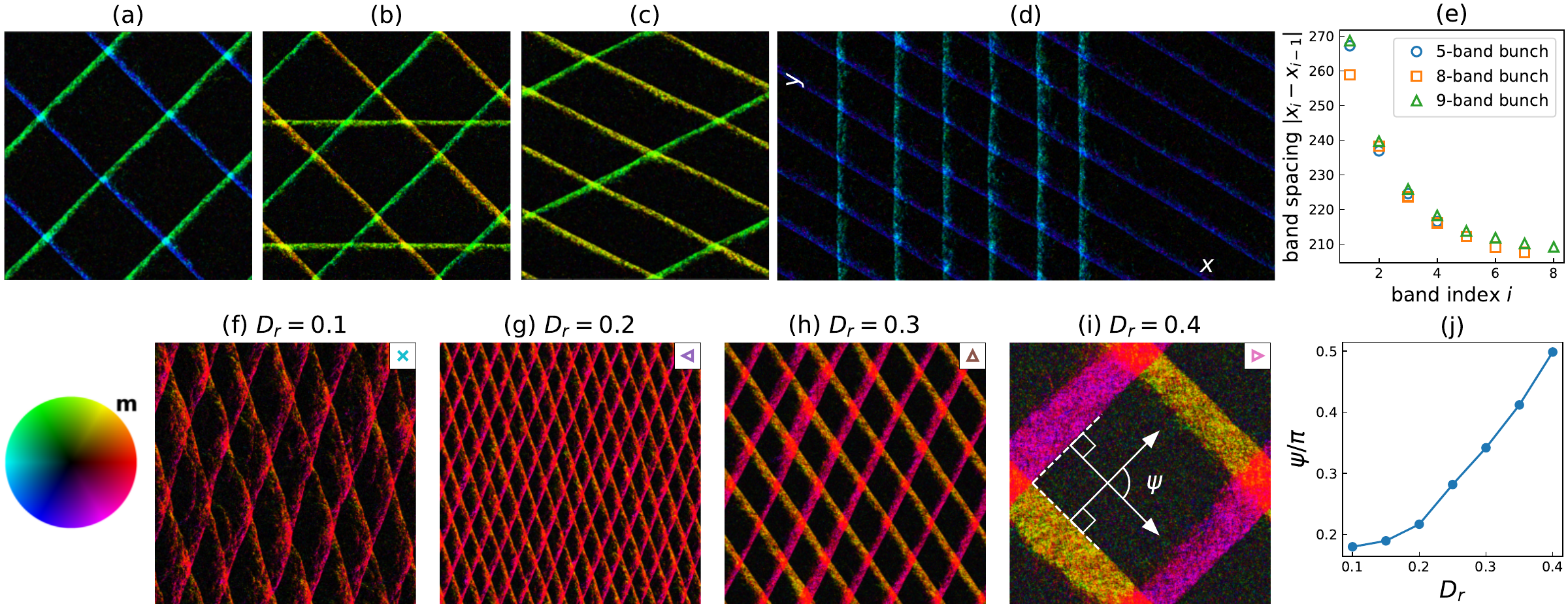}
    \caption{Stable configurations with crossing bands. 
(a-c) snapshots of momentum fields of various stable configurations with crossing bands obtained near the gas binodal 
(discrete-time VM with $v_0=0.5$, $\rho_0=1$, $L_x=L_y=2400$, (a) $\eta=0.39$, (b,c) $\eta=0.38$)
(d,e) bunching of thin bands in unbalanced cross-sea configurations (discrete-time VM with $v_0=1$,
    $\eta=0.36$, $\rho_0=0.637$).
    (d) snapshot of momentum field ($L_x=2L_y=2400$): a single diagonal band wraps around the torus while 6 vertical bands traveling to the left bunched together after transient.
    (e) spacing between two consecutive bands vs band index in bunches of thin bands similar to that shown in (d) (band index is counted from the leading band of the bunch). 
    Here, a large system of size $L_x=8 L_y=9600$ is studied, in which the three bunches, consisting of 5, 8 and 9 bands respectively, can spatially coexist even up to $t=5\times 10^7$.
    (f-i) snapshots of momentum field in balanced regular cross-sea
      configurations observed near and above the bands binodal (symbols in the top right corner correspond to those on the phase diagram in Fig.~\ref{fig8}(a)) (continuous-time VM, $L_x=L_y=2400$, $D_r=0.1$, 0.2, 0.3, 0.4 and $\rho_0=0.36$, $1.3$, $2$, $4.3$ from left to right).
      (j)  variation with noise strength $D_r$ of the crossing angle $\Psi$, defined as in panel (i) for configurations similar to those shown in (f-i). 
        }
\label{fcross}
\end{figure*}

\subsubsection{Inverse bands}
\label{IB}

At strong noise but larger densities than those considered above, 
one reaches a regime where there is so little space between fat bands 
that their interfaces interact strongly across what become narrow strips of gas with well-defined profiles
(Fig.~\ref{figIBs}(a,b)). 
The lower panels of Fig.~\ref{figIBs}(a) show that the density and order time-averaged profiles 
of these ``inverse bands'' (in blue)
are directly ``inherited'' from those of the fat bands found at the same noise strength but lower density (in orange).
Note, however, that order does not reach zero at the ``bottom'' of the inverse bands, which means that 
they have a small but finite speed at which they move in the same direction as the TT liquid surrounding them.

As mentioned in the SP, 
it is tempting to see these inverse bands as the microphases of some ``reverse'' phase separation scenario. 
In small-ish systems such as those shown there ($L_\|=1200$, $L_\perp=300$), one observes equispaced 
inverse bands whose number decreases with density, but not linearly (see Fig.~2(b,e) in  \cite{SP}).
However, in longer systems such as the $L_\|=9600$ one shown in Fig.~\ref{figIBs}(b) and in SMov4~\cite{SUPP}, 
these equispaced configurations are not stable, and the number and location of inverse bands present varies, albeit on very long time scales. In Fig.~\ref{figIBs}(c), we show how the number of inverse bands become more and more unsteady as the length of the system is increased. 
The space-time plot in Fig.~\ref{figIBs}(d) reveals interactions across the trains of inverse bands, 
and propagative modes mediating this instability.

Finally, all the above results were obtained with relatively moderate values of $L_\perp$. 
For much wider systems, as for fat bands, we observe what looks like a long-wavelength instability 
of inverse bands (Fig.~\ref{figIBs}(e) and SMov5~\cite{SUPP}). 
Contrary to what we saw with fat bands, though, this instability of long inverse bands does not lead to the
emergence of thin bands. Instead, a ``chaos of inverse bands'' sets in.

\subsection{Crossing bands}

As mentioned in the introduction, in \cite{kursten2020dry} K\"ursten and Ihle put forward a bona fide new phase
made of regular configurations of crossing bands, an example of which is shown in Fig.~\ref{fig1}(d).
These cross-sea patterns, 
they argue, constitute an independent phase because they are not just the simple 
superposition of two smectic trains of thin bands: the density observed at the crossing points is much higher than twice the maximum density of the bands elsewhere, and their crossing angle is ``strongly selected'', 
taking values near $\tfrac{\pi}{4}$.

Here we present results leading us to conclude that a better description of the situation is in terms of a 
single phase within which various band patterns coexist and compete. 
They were obtained using either the discrete-time or the continuous-time VM. 
All reported phenomena, though, are observed in both versions.

Let us note first that cross-sea configurations not being a linear superposition of two smectic patterns of bands 
is by no means surprising given that we are here in presence of nonlinear waves solutions of a nonlinear system. 
Band crossing points are indeed very high density steep structures that play a major role in the 
complicated dynamics inside and between the bands involved.

In the Supplemental Material of their paper, K\"ursten and Ihle report
the coexistence of a variety of crossing band patterns in the same system but following different random initial conditions, some of them found rather far away from the region where regular cross-sea patterns dominate. 
We show now that we are able to generate stable configurations where bands cross each other 
almost anywhere in the coexistence region.

Using ``engineered" initial conditions carefully constructed from portions of bands, 
allows us to obtain stable configurations almost at will.
Examples involving only standard thin bands are shown in Fig.~\ref{fcross}(a-c).
(Details about ``band engineering'' can be found in Appendix~\ref{appBE}.)
Importantly, this can be achieved even arbitrarily close to the gas binodal line, where we can observe, e.g., stable
configurations with only two bands crossing each other. Note also that we can engineer configurations with 
bands crossing at various angles, including right angles, 
eroding the angle selection put forward in \cite{kursten2020dry}. 
We were even able to create stable ``triple" cross-sea configurations in which three sets of parallel bands travel in different directions.

The above diversity of patterns, however, decreases when increasing density or decreasing noise strength.
In particular, as argued in \cite{chate2020dry} and observed by K\"ursten and Ihle in  \cite{kursten2020dry}, 
smectic configurations gradually disappear: the microphase separation scenario,
 in which a unique band profile is selected at a given noise strength \cite{solon2015phase},
implies that there is then not enough space to arrange the liquid fraction into a smectic of bands, and only 
configurations with crossing bands, which allow to fit in a larger total band length, are possible.

Rather frequently observed  then are ``unbalanced'' cross-sea configurations in which the numbers of bands 
present in each parallel set differ. 
Interestingly, when one of these sets is parallel to one axis of the spatial domain, 
we observe systematically that bands in this set eventually evolve to a final arrangement 
in which they are bunched together   (Fig.~\ref{fcross}(d)). 
While the inter-band spacing of the other (oblique) set is regular, 
that of the bunched bands decreases from the leading edge to the tail of the bunch (Fig.~\ref{fcross}(e) and SMov6~\cite{SUPP}).
This indicates that crossing band configurations are not just ruled by band intersection points, since the geometry of the spatial domain in which they travel plays a role.

\begin{figure*}[t!]
  \includegraphics[width=\textwidth]{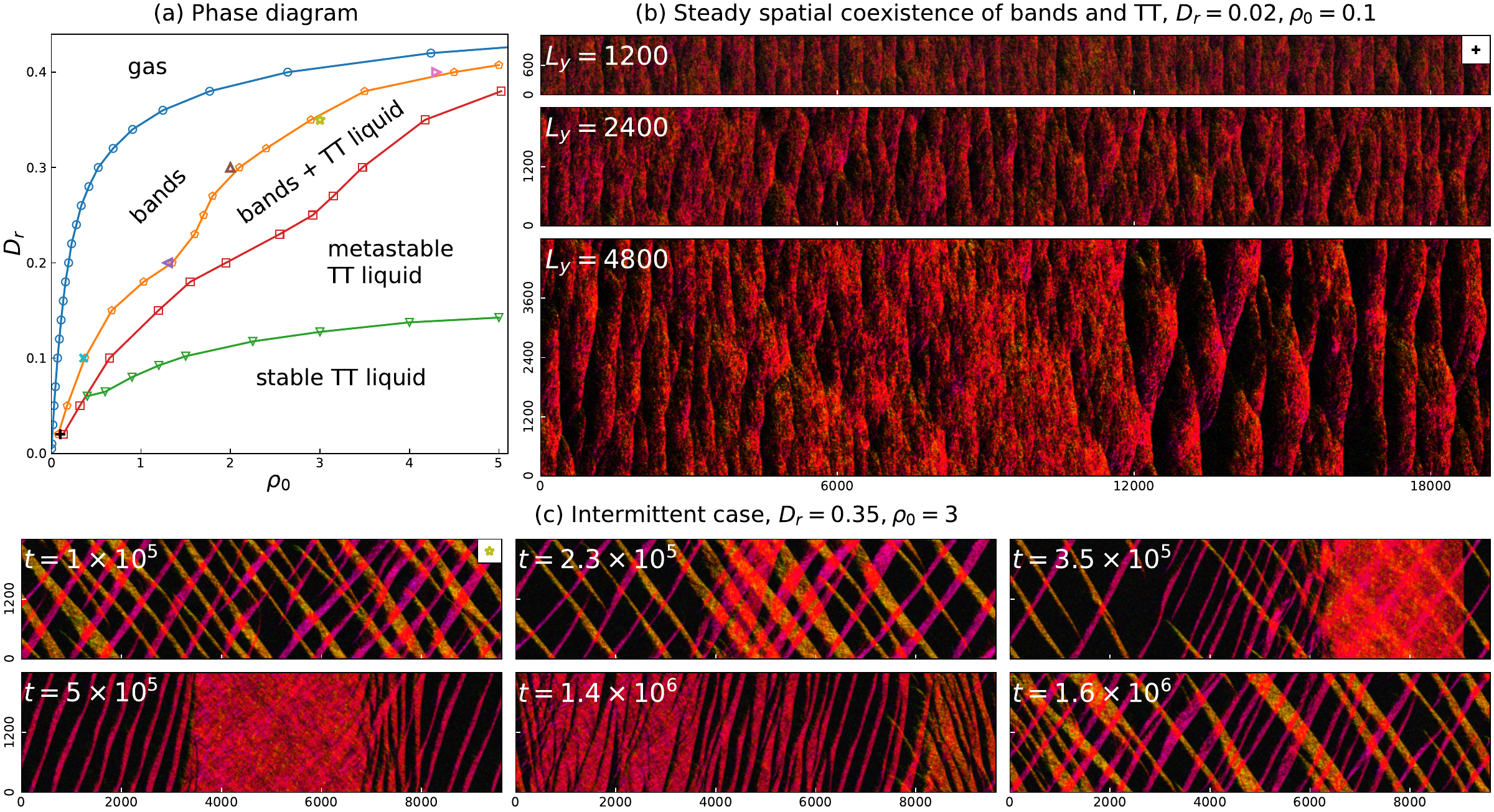}
    \caption{Nature of the liquid binodal in the low noise and high-noise regimes.
      (a) new phase diagram as elaborated in the SP.
      (b) snapshots of momentum field taken in 3 long systems ($L_x=9600$) of increasing width at very weak noise ($D_r=0.02$) and density $\rho_0=0.1$ chosen near and above the bands binodal. Only the widest system ($L_y =4800$) starts exhibiting a cross-sea pattern coexisting with TT liquid.
      (c) momentum field snapshots taken at various times in a system showing intermittent super-phase separation at strong noise ($D_r=0.35$, $\rho_0=3$, $L_x=9600$, $L_y=2400$).
      	}
\label{fig8}
\end{figure*}

Regular, balanced, cross-sea patterns only dominate in a thin strip of parameter space bordering the 
bands binodal in the new phase diagram built in the SP (see also the next section, where this phase diagram is
reproduced in Fig.~\ref{fig8}(a), and the bands binodal is the orange line). It is only in this narrow region, at fixed parameters, 
that the crossing angle of bands is rather sharply selected. 
However, moving along this strip, this well-defined crossing angle $\Psi$ 
increases continuously as the noise strength increases (Fig.~\ref{fcross}(f-j)). 
It tends to zero in the vanishing noise limit, and to $\tfrac{\pi}{2}$ in the infinite-density limit.

\subsection{Coexisting stable configurations?}

All the band configurations described above are ``stable'' in the sense that we can observe them 
for tens if not hundreds of millions of timesteps (the typical duration of our runs). 
Many of them coexist in the same system at the same parameters over these timescales. 

If we were discussing the solutions of a deterministic dynamical system, this would be the signature of the 
existence of multiple attractors with disjoint basins of attraction. 
But the Vicsek model is ruled by stochastic equations, and its intrinsic fluctuations mean that probabilities to 
transit from one stable configuration to another, although minuscule, cannot vanish. 
Some configurations must thus be more stable than others, 
something that could be quantified by comparing their typical lifetimes. 
Short of being able to do this systematically, we can nevertheless propose the following picture
for the weak noise regime where stable thin band configurations dominate.

As we have seen, in a large region bordering the gas binodal, multiple configurations with crossing bands coexist with 
smectic arrangements. These latter arrangements, we believe, are the most stable ---in the sense defined above--- because 
the band crossing points are highly localized and fluctuating structures more able to significantly perturb the system locally and lead to band re-arrangements. 
At large densities, however, there is too much mass above the gas level (set by the noise strength) to distribute it
in a smectic configuration and configurations with crossing bands then dominate. 

\section{Building the new phase diagram}
\label{diagram}

The accompanying short paper \cite{SP} is devoted to establishing a phase diagram updated from
the ``standard one'' of Fig.~\ref{fig1}(a). 
This new phase diagram, reproduced here in Fig.~\ref{fig8}(a) for convenience, derives from the knowledge
acquired by studying the VM at length- and time-scales never explored before. 
Note that it ignores largely the diversity of band patterns described in Section~\ref{diversity}, many of which
were found to be unstable at very large scales.

In addition to the green line marking the limit of metastability of the Toner-Tu liquid, the main change is the splitting
of the coexistence region in two. The new orange ``bands binodal line" runs in between the blue gas binodal
and the red liquid binodal. These orange and red lines are defined as the binodals of the ``super phase separation''
scenario described in the SP, in which band patterns, themselves the result of the initial phase separation,
coexist in space with a large domain of Toner-Tu liquid.

Below, we provide further details about the liquid binodal and the new bands binodal
(respectively the red and orange lines  in Fig.~\ref{fig8}(a))
that were not given in the SP due to space constraints. 
We then explore the metastability of some of the emblematic band patterns of the coexistence phase,
beyond the study presented in the SP for the TT ordered liquid, which led to the determination of 
green line in Fig.~\ref{fig8}(a).

\begin{figure*}[t!]
  \includegraphics[width=\textwidth]{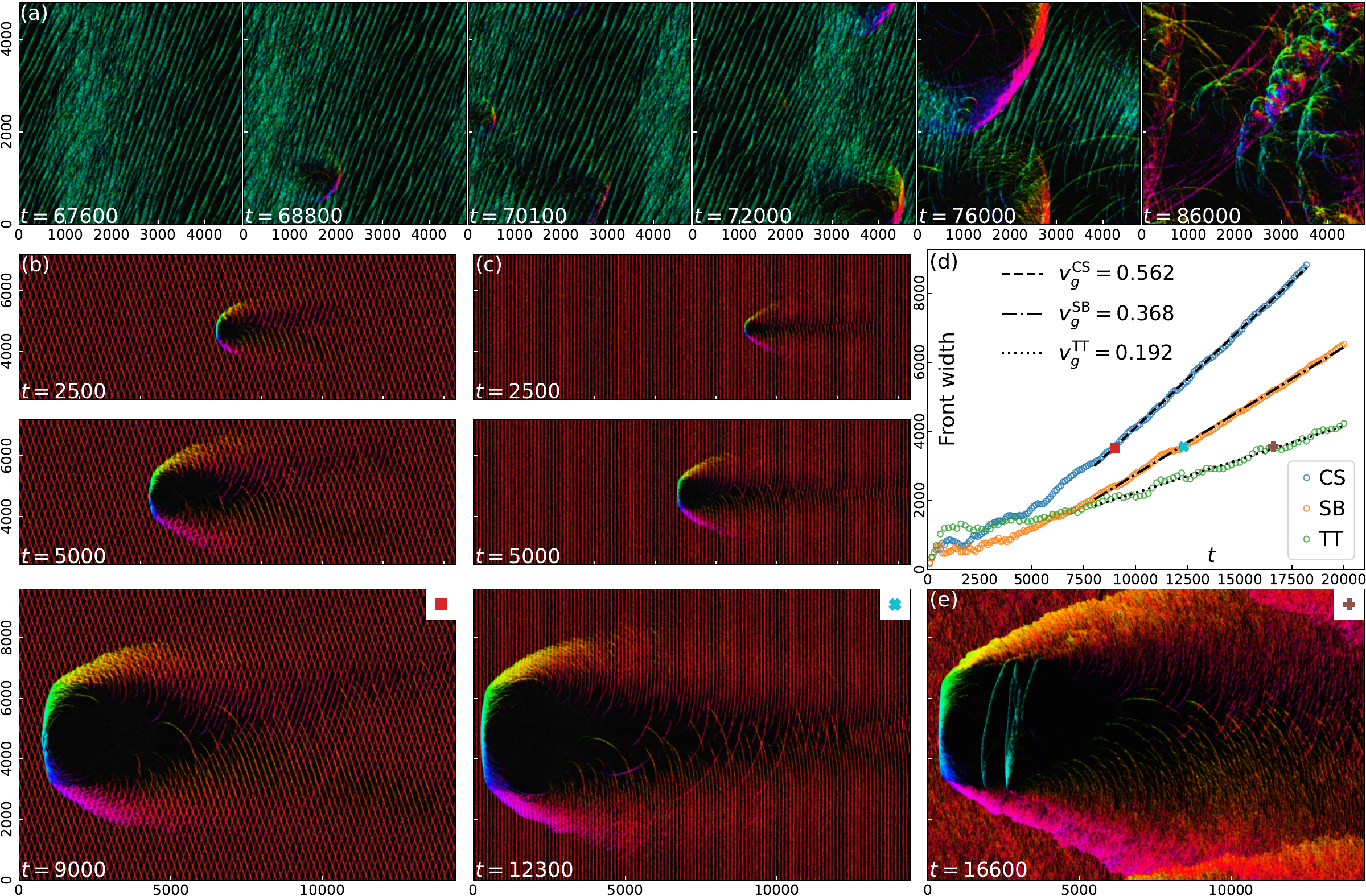}
    \caption{Metastability in the coexistence phase.
    (a) snapshots of momentum field during spontaneous nucleation and growth of counter propagating droplets
    (discrete-time VM, $\eta=0.43$, $\rho_0=1.4$, $v_0=1$, $L_x=L_y=4800$)
    (b,c) snapshots of momentum field after the introduction of a counter-going perturbation in regular cross-sea (b) and smectic (c) arrangements of bands (continuous-time dynamics with same parameters for (b) and (c): $D_r=0.2$, $\rho_0=0.8$, $v_0=1$, $L_x=38400$, $L_y=9600$).
    (d) width of growing front for the cases shown in (b,c), plus a case of growth against the TT liquid present at same parameters except $D_r=0.11$, a snapshot of which is shown in (e). The ballistic regime sets in before $t=10000$, after which the growth speed $v_g$ is measured from the linear fits shown by the black lines.
    Bottom panels (including (e)): snapshots of momentum field for the 3 cases, taken in the ballistic growth regime when the front widths have approximately the same size ($\sim 3500$). Symbols in the top-right corner correspond to those in (d).	
    }
\label{fig9} 
\end{figure*}

\subsection{The liquid and bands binodals}

Previous numerical work, performed on systems of moderate size, used ad-hoc quantities to 
define and locate the liquid binodal separating the coexistence region from the Toner-Tu liquid.
Moreover, only the low-noise low-density region of parameter space, where thin bands dominate, was considered.
References \cite{chate2008collective} and \cite{solon2015phase}, 
in particular, used the variances in space
of density and order profiles calculated along the direction of global order, averaging over the transverse direction. 
These variances are small in the Toner-Tu liquid region, and 
take finite values when the system displays a smectic arrangement of bands.
But this signature of the transition gets blurred in configurations with large transverse dimension,
in particular in the presence of cross-sea configurations, which dominate in this region.

In the SP, we show that the liquid binodal is unambiguously defined in large-enough systems 
where bands and the Toner-Tu liquid coexist {\it in space}. 
Then what we coined a ``super phase separation'' scenario takes place in which the Toner-Tu liquid and bands
(which themselves arise from microphase separation) coexist in space. 
The orange and red lines in Fig.~\ref{fig8}(a) are the binodals of this super phase separation.

Here we discuss further the band patterns coexisting with the Toner-Tu liquid between these two lines.
When coexistence in space is steady, i.e. at not too strong noise, the Toner-Tu liquid domain evolves 
within a regular cross-sea pattern, eating it up slowly, since it moves faster (cf. Fig.~3(a) in the SP). 
It sheds particles at its back that quickly reform the cross-sea pattern. 

Decreasing the noise strength, individual bands become wider (see Section~\ref{twolimits}),
and they cross at a smaller angle in the cross-sea pattern (Fig.~\ref{fcross}(f-j)). 
This imposes to consider larger and larger systems in the transverse direction, otherwise spatial coexistence
disappears (Fig.~\ref{fig8}(b)  and SMov7~\cite{SUPP}). The low-noise low-density limit is thus difficult to approach numerically.

As shown in the SP, with strong noise the super phase separation becomes intermittent in time. 
The Toner-Tu liquid domain disappears and reforms, an irregular dynamics that takes place on long timescales.
The liquid domain emerges from some instability of a more or less regular cross-sea pattern. In Fig.~\ref{fig8}(c) and in SMov8~\cite{SUPP},
we show typical snapshots taken along a run performed in this intermittent regime, but at a noise level much higher than that shown in the SP ($D_r=0.35$ instead of $0.23$). Whereas the general features of this regime remain the same, there are some clear differences. The cross-sea pattern that breaks and reforms at high noise is now
made of fat bands less regularly arranged than in Fig.~3(e) in the SP. 
When broken, smectic regions appear, also made of fat bands.

As we have seen above (Fig.~\ref{figFBs}(c,d)), fat bands are unstable when long enough. 
This implies that the complicated dynamics present in these intermittent phase separation regimes, 
although observed on very large systems comprising hundreds of millions of particles, 
probably do not represent the asymptotic regime.

\subsection{Metastability}
\label{meta}

Almost all the configurations presented so far possess global polar order. 
It has emerged recently that globally ordered phases in active matter can be metastable. 
Constant density flocks have been shown to be always metastable, susceptible to the nucleation 
of aster defects that 
grow and then proliferate to leave an active foam \cite{besse2022metastability}. 
Similarly, the ordered phase of the active Ising model was shown to be always metastable, 
with counter-going droplets that grow ballistically \cite{benvegnen2023metastability}. 
For the Vicsek model, a similar observation was actually made first by Codina et al. for the TT liquid, but
with metastability present only for strong enough noise \cite{codina2022small}. 
The mechanism underlying this excitability transition was recently elucidated in \cite{granek2026how}.

The green line in Fig.~\ref{fig8}(a), determined in the SP, 
marks the weak noise metastability limit of the Toner-Tu ordered liquid.
What about a strong-noise limit? The TT liquid is metastable between the 
green line and the red liquid binodal, but what about stronger noise regimes in the coexistence phase?
Given the diversity and complexity of the band patterns described above, 
a comprehensive answer to this question is a  task beyond the scope of this paper. 
But partial information can be gained that we present below.

We first report here the observation of the {\it spontaneous} breakdown of order in regimes showing 
fat and inverse bands (Fig.~\ref{fig9}(a) and SMov9~\cite{SUPP}). In such rare events,
a fluctuation nucleates a counter-going, growing front that quickly destroys current order, often leading to 
a reversal at longer times, and to some large-scale band chaos in very large systems.

In practice, however, observing such spontaneous breakdown of order in reasonable time and system size 
is limited to rather special regimes. Beyond these, one has to resort to the methodology used in
\cite{codina2022small,benvegnen2023metastability} and in the SP and
manually introduce a localized group of particles oriented against the mean flow and then observe whether it  
grows into an invading counter-going droplet. As shown in the SP for the TT liquid, 
the shape and density of this group of particles
do not matter much. Of course it has to contain enough particles, and this number varies with the shape of the small domain in which they are introduced, but beyond this, once the ballistic growth is triggered, it occurs with
the same growth speed $v_g$, and the growing droplets then have identical shapes. 
These observations hold when studying the metastability of the band patterns of the coexistence region.

We considered mostly the two emblematic cases of the coexistence phase,
a regular smectic arrangement and a balanced cross-sea pattern. We used the same type of initial perturbation
shown to be efficient in the SP, namely a transverse $32\times5$ rectangular domain of density $\rho_d$.
In Fig.~\ref{fig9}(b,c), we show snapshots of the evolution of this initial perturbation in successful cases, where 
it eventually reaches a ballistic growth regime. 
Interestingly, the two background band patterns were engineered to be at the same parameters, 
so that any difference must be traced back to the geometry of these patterns.
One obvious difference is that $v_g$ is larger in the cross-sea than in the smectic pattern. 
This is confirmed by the time series of the front width shown in Fig.~\ref{fig9}(d),
with corresponding growth dynamics showing in SMov10~\cite{SUPP}.
On the other hand, the shape of the growing droplet, 
once it has reached the ballistic regime, is surprisingly similar in both cases, and
very close to that observed in the TT liquid. In the cross-sea case, the leading front is slightly more irregular 
and more curved than in the other two cases (Fig.~\ref{fig9}(e)).

We have observed that cross-sea and smectic patterns other than the two presented above are also metastable. 
This leads to wonder whether possibly all band patterns are eventually metastable. 
Of course, near the gas binodal, where solutions comprise very few bands separated by arbitrarily large gas areas,
the particles of a counter-going droplet would have time to spread, weakening the front they form, which can then be overcome by the next incoming band. Whether such a droplet could succeed to grow forever in such a case is 
a complicated question involving many times- and length-scales that we leave for future studies.

As a matter of fact, it is still not known whether the single thin band solution observed near the gas binodal is 
really stable. The full 2D linear stability analysis of such a solution of the Toner-Tu equations have not been performed
yet. If it were found (against all available numerical evidence) that such an instability is present, then the whole
coexistence region would be one, asymptotically, of very large scale ``band chaos''.

\section{Discussion and perspectives}
\label{discussion}

\subsection{Summary}

We have found that the coexistence phase of the two-dimensional 
Vicsek model is more complex than the simple 
microphase separation picture of smectic arrangements of traveling thin bands put forward in \cite{solon2015phase}. 
While we have argued that the cross-sea configurations studied in \cite{kursten2020dry} do {\it not} constitute 
a bona fide sub-phase, we have uncovered a number of other large-scale patterns.
We conclude, though, that most of them are unstable and disappear in the limit of large transverse size,
to be replaced by the standard thin bands. 
This is the case of the lanes of counter-going trains of bands. This is also true of
 the fat bands that may deceptively suggest a macrophase separation scenario at strong noise and high density. 
The inverse bands present in this latter regime close to the liquid binodal line 
also undergo a large-wavelength instability, with the subsequent regime being a globally-ordered but chaotic one 
in which bits of inverse bands evolve irregularly. 
We have also started exploring whether some band configurations are metastable. 
We found that both regular smectic and cross-sea patterns can be destroyed by counter-going droplets
very similar to those observed in the Toner-Tu liquid. 

\subsection{Open questions}

To be true, we still ignore whether the standard thin bands are stable objects. 
Simulations of a single such band in very large systems (something possible near the gas binodal)
do not reveal any long wavelength instability. 
Even if linearly stable, such a long band nevertheless constitutes a propagating, fluctuating  one-dimensional interface~\cite{kardar1986dynamic}. 
Even if extremely weak, the roughening properties of a thin band and band interactions have the potential
to impact the stability of the various configurations at play in the coexistence phase.
Indeed, smectic order could be destroyed in trains of very long bands, possibly ruining also global polar order~\cite{chen2013universality}. 
In such a scenario, that part of the coexistence phase would actually constitute a disordered phase.

Note that a similar conclusion would be reached if, as discussed in the previous section, 
the entire coexistence phase were metastable. 

\subsection{Continuous theories}

We finally discuss to what extent continuous theories~\cite{marchetti2013hydrodynamics,bertin2009hydrodynamic,mishra2010fluctuations,gopinath2012dynamical,ihle2013invasion,solon2015pattern} can account for the phenomena described here at 
particle level in the Vicsek model.

Toner-Tu like equations derived from the VM via a Boltzmann kinetic equation are known to be able to 
reproduce qualitatively the basic phase diagram shown in Fig.~\ref{fig1}(a)~\cite{bertin2009hydrodynamic,peshkov2014boltzmann}. 
The homogeneous ordered solution is generically linearly unstable in a parameter domain bordering the onset of order.
This domain defines, in the phase separation language, the spinodal lines. 
Inside this domain, but also rather far outside,
one finds, at fixed parameters, a multitude of coexisting smectic band pattern solutions. 
A fraction of them is found to be linearly stable in systems of modest transverse extent~\cite{solon2015pattern}. 
Simulations of stochastic Toner-Tu like equations revealed that a single solution is selected by fluctuations~\cite{solon2015phase}. 
The binodal lines defining the coexistence phase are understood as the limits of stability of the selected solution.

These results only deal with smectic trains of bands. Simulations of Toner-Tu equations have not been shown, 
so far, to display any of the other patterns reported here. 
On the other hand, as mentioned earlier, 
counter-going lanes similar to those studied in Section~\ref{lanes}, have been reported in 
simulations of a Boltzmann equation derived from a Vicsek model \cite{thuroff2014numerical}.
In their study of the metastability of the TT liquid, Codina {\it et al.} \cite{codina2022small}
found that truncations of the 
Boltzmann equation keeping at least 4 fields are able to account for the growth of counter-going droplet. 

That the kinetic level is able to account for complicated patterns like counter-going lanes or the fronts
of the ballistically growing droplets in the metastability region indicates that
different hydrodynamic-level transport coefficients and/or higher-order terms might be enough to have Toner-Tu like 
equations sustaining the many patterns reported here. 

A similar conclusion can be reached when considering the fat and inverse band solutions found to be stable 
in systems of not-so-long transverse dimension (Figs.~\ref{figFBs},\ref{figIBs}). Indeed, as we argue in Appendix~\ref{appFB},
their density and order profiles are incompatible 
with the simple dynamical system framework developed from the TT equations that was studied in \cite{solon2015pattern,caussin2014emergent}.

Searching for continuous theories able to account for the phenomenology presented here is an important 
endeavor left for future studies, together with the key pending questions discussed above.
More generally, we stress again that there are at least two good reasons
to keep studying the Vicsek model and in particular its coexistence phase: (i) it is important to know 
all the properties of such a `central' model, irrespective of its weak direct relevance to real situations; 
(ii) some of the pending questions listed above constitute an appealing playground
 for exploring and testing new out-of-equilibrium phenomena.
 
\acknowledgments
We thank Alexandre Solon and Yongfeng Zhao for a careful reading of the manuscript.
This work is supported by the National Natural Science Foundation of China 
(Grants No.~12504239 to Y. D. and No.~12275188 to X.-q.S.); the Basic Research Program of Jiangsu (No.~BK20250788 to Y. D.); and the Natural Science Foundation of Jiangsu Province (No. BK20233001).

\appendix

\section{Band engineering}
\label{appBE}

Configurations made of standard thin bands, typically observed in the weak noise regime, have well-known 
properties stemming from the microphase separation scenario put forward in \cite{solon2015phase}.
In short: at a given noise strength, there exists a unique band profile and a well-defined gas density, 
both independent of the global density. This allows to engineer many different band configurations.

\begin{figure}[t!]
  \includegraphics[width=\columnwidth]{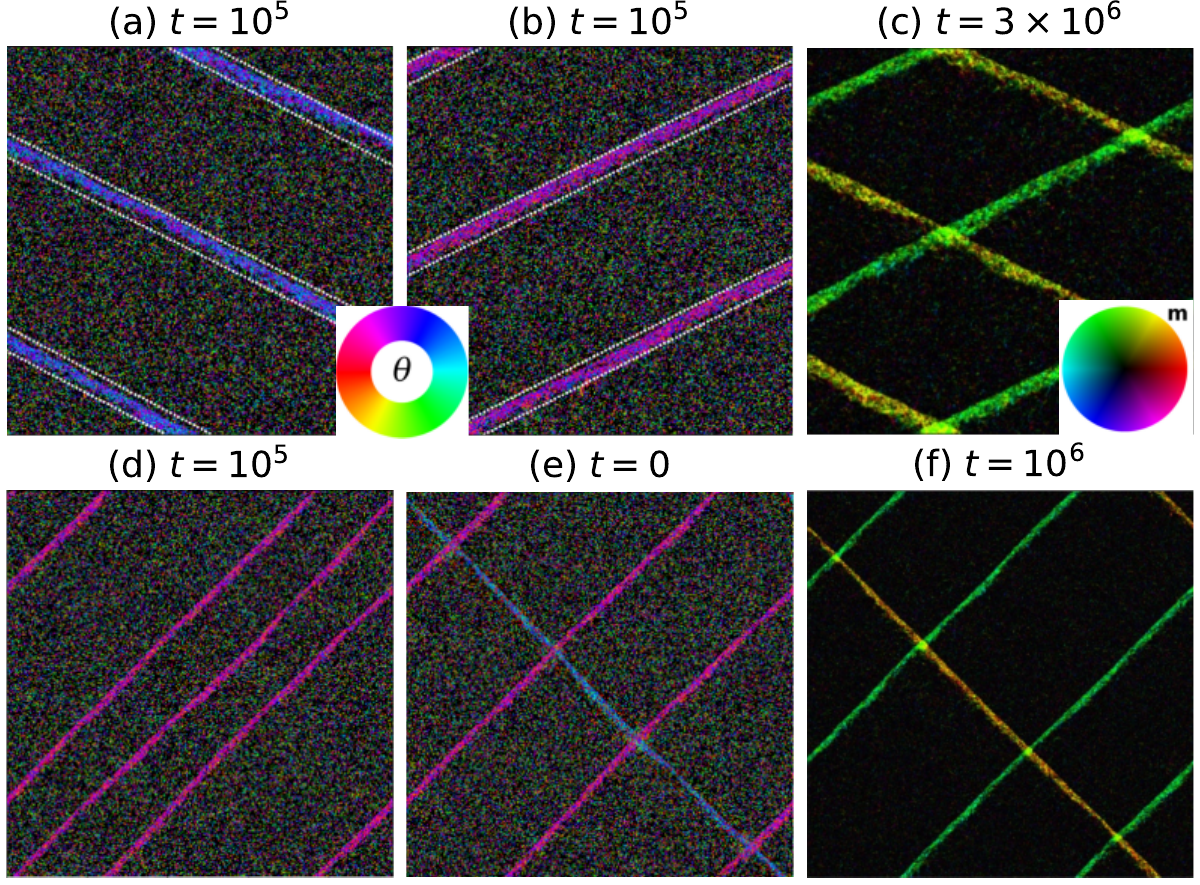}
    \caption{Engineering band configurations. (Discrete-time VM with $v_0=1$, $\eta=0.43$, $\rho_0=0.637$ in all panels, except (c) where $\rho_0=0.836$.) 
    Top row: superposition of two configurations (a,b) obtained from different initial conditions, leading to the stable configuration shown in (c) ($1200\times 1200$ domain).
    Bottom row: the band present in the diagonal in (d), was rotated by $\tfrac{\pi}{2}$, leading to the initial configuration in (e), shown in (f) to be stable ($2400\times 2400$ domain).
     }
\label{figABE} 
\end{figure}

In Fig.~\ref{figABE}(a-c), we show how two smectic configurations obtained in the same system
from different initial conditions can be combined to create a stable cross-sea pattern. 
(Note that this can be done arbitrarily close to the gas binodal.)
The bands present in the two systems are first extracted using a density threshold at $1.5\rho_{gas}$, then copied in a new system, after which particles are introduced randomly in the space surrounding the bands until the gas density is reached. The obtained configuration (c) is stable. Note that this neglects the strong density peaks observed usually at band crossings, but this can also be taken into account gradually by injecting more particles in the gas once these 
crossings have taken their steady structure.   

A simpler procedure is illustrated in Fig.~\ref{figABE}(d-f), where one band is extracted from a given smectic configuration, rotated by $\tfrac{\pi}{2}$, and re-introduced in the system. 
This typically does not even require gas density adjustments, and thus conserves the global density. 

\section{Incompatibility of fat and inverse band profiles with standard Toner-Tu equations}
\label{appFB}

\begin{figure}[b!]
  \includegraphics[width=\columnwidth]{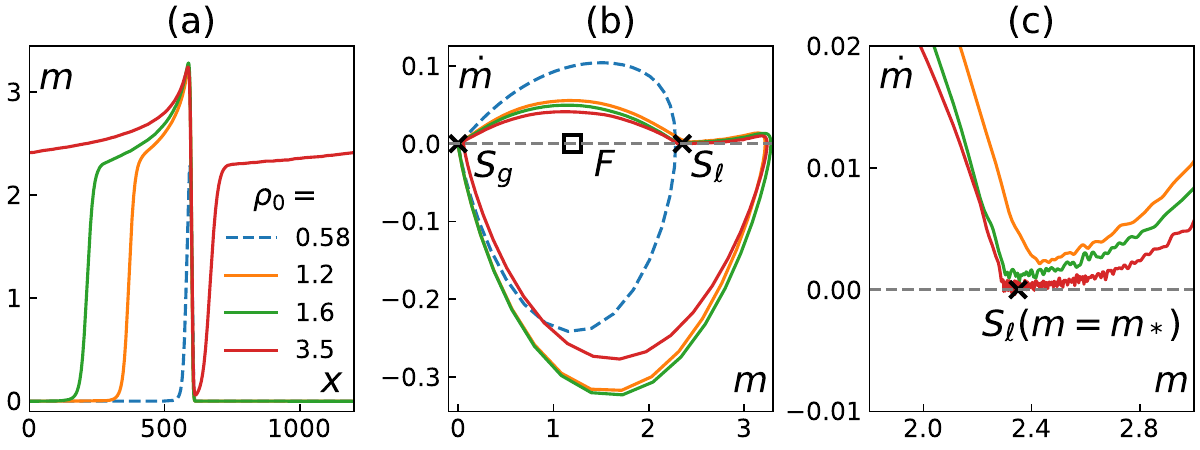}
    \caption{Incompatibility of fat and inverse bands with standard Toner-Tu equations.
    (a) time- and $y$-averaged momentum profiles of fat and inverse bands moving along $x$ 
    (Continuous-time VM with $v_0=1$, $D_r=0.3$.) 
    (b) Corresponding orbits in the $(m,\dot{m})$ plane, plus the orbit (dashed blue line)
    of a standard thin band smectic configuration obtained at $\rho_0=0.58$. 
    (c) Zoom of the fat and inverse orbits shown in (b) in the vicinity of the liquid saddle $S_\ell$.
    The positions of $F$ and $S_\ell$ are schematic, illustrating the fixed-point structure of the standard TT equations rather than fixed points explicitly determined from the measured profiles.
    }
\label{figAFB} 
\end{figure}

In \cite{solon2015pattern,caussin2014emergent}, it was demonstrated that the standard Toner-Tu equations do support multiple
solutions of smectic arrangements of traveling bands. 
This was achieved via a mapping to a simple dynamical system, looking for solutions homogeneous in the transverse
direction, traveling at a constant speed $c$. Assuming the global motion is along the $x$ axis, the TT equations
can be expressed in the frame moving at speed $c$, as ordinary differential equations for the reduced variable 
$z=x-ct$, which take the form of the two-dimensional Li{\'e}nard system~\cite{gine2017lienard}:
\begin{equation}
\ddot{m} = A(m) + B(m) \dot{m}
\end{equation}
with the friction term $B(m)$  linear in m, and the Landau-Ginzburg term $A(m) = a_2 m + a_3 m^2 + a_4 m^3$.
This system is organized by a three-fixed-point spatial dynamics, $S_g~-~F~-~S_\ell$, where 
$S_g$ and $S_\ell$ are saddle-type fixed points associated with gas- and liquid-side homogeneous states
and $F$ is a positive-index interior fixed point.
With periodic boundary conditions, the smectic
band configurations correspond to limit cycles that can come very close to $S_g$ (see, e.g. the dashed blue orbit in 
Fig.~\ref{figAFB}(b)), while macrophase separation is
organized by heteroclinic connections between the gas and liquid saddles.

The typical profiles of fat and inverse bands shown in Fig.~\ref{figAFB}(a), 
represented in the $(m,\dot{m})$ plane in 
Fig.~\ref{figAFB}(b), are incompatible with this 3-fixed points structure.
While standard thin bands have the familiar single-upper-hump/single-lower-valley
geometry in this plane (blue orbit), 
fat bands and inverse bands, in contrast, show:
(i) an upper branch $\dot{m}_+(m) > 0$ with two maxima, the left one higher than the right one;
(ii) a shallow upper valley (UV) connecting the two maxima;
(iii) a lower branch $\dot{m}_-(m) < 0$ with one comparatively deep minimum.

As $\rho_0$ increases, the upper-valley value $\dot{m}_{UV}$ decreases and the bottleneck approaches the $\dot{m}=0$ axis (Fig.~\ref{figAFB}(c)).
Since the UV is a local extremum of the upper branch, $d\dot{m}_+/dm=0$, implying $\ddot{m}=0$ at this point.
If simultaneously $\dot{m}_{UV}\to 0$, the bottleneck approaches a spatial fixed point at an intermediate $m=m_*$ within the range spanned by the orbit.
This suggests a fixed-point structure beyond the standard $S_g-F-S_\ell$ picture, in which periodic orbits enclose only $F$, while the two saddles lie outside~\cite{solon2015pattern}.
Indeed, if a periodic orbit were to enclose both $F$ (index $+1$) and $S_\ell$ (index $-1$), the Poincar\'e index theorem would require at least one additional positive-index fixed point, since the total enclosed index must be $+1$~\cite{strogatzNonlinearDynamicsChaos2024}.
This would require at least four fixed points, beyond what the standard cubic Landau-Ginzburg term $A(m)$ can support.
The observed bottleneck geometry thus suggests an incompatibility of fat and inverse band profiles with the standard TT equations~\cite{solon2015pattern,caussin2014emergent}.

\section{Description of the Supplemental Movies}
\begin{itemize}
    \item \texttt{SMov1.mp4} corresponds to Fig.~\ref{fig_short_lane}(d, e) and illustrates a wide system of size $(L_x=9600, L_y=2400)$ with narrow lane bunching.
    \item \texttt{SMov2.mp4} corresponds to Fig.~\ref{fig_long_lane}(b), showing that in a system of size $(L_x=9600, L_y=4800)$, the initial narrow lanes become unstable quickly, leaving two wide lanes that survive after a long run.
    \item \texttt{SMov3.mp4} corresponds to Fig.~\ref{figFBs}(c), showing the momentum field of a long fat band, which becomes unstable and eventually superseded by a cross-sea configuration.
    \item \texttt{SMov4.mp4} corresponds to the bottom panel of Fig.~\ref{figIBs}(b), showing that equally spaced inverse bands are not stable in a long system with $L_\parallel=9600$.  
    \item \texttt{SMov5.mp4} corresponds to Fig.~\ref{figIBs}(e), where a long inverse band undergoes transverse instability, leaving a chaos of inverse bands.
     \item \texttt{SMov6.mp4} corresponds to Fig.~\ref{fcross}(e), showing three bunches of vertical bands that can spatially coexist in long time simulations. The same parameters as Fig.~\ref{fcross}(d) but with much larger $L_x=9600$.
     \item \texttt{SMov7.mp4} corresponds to the bottom panel of Fig.~\ref{fig8}(b), where the steady super-phase separation arises in a system with sufficiently large transverse size $(L_y=4800)$.
     \item \texttt{SMov8.mp4} corresponds to Fig.~\ref{fig8}(c), showing intermittent super-phase separation in the strong-noise regime.
     \item \texttt{SMov9.mp4} corresponds to Fig.~\ref{fig9}(a), where spontaneous nucleation and growth of counter-propagating droplets are observed.
     \item \texttt{SMov10.mp4} corresponds to Fig.~\ref{fig9}(b-e), showing the growth of initial counter-propagating perturbations in the background of cross sea, smectic bands and Toner-Tu liquid, respectively.
\end{itemize}

\bibliography{Biblio-current.bib}

\end{document}